\documentclass{aa}

\usepackage{graphics}
\begin{document}
\thesaurus{05(08.16.4;
08.13.2;  
10.03.1;
10.19.2; 
13.09.6
)}

\title{ISOGAL-DENIS detection of red giants with weak mass loss in the Galactic
Bulge\thanks{This is paper no. 4 in a refereed journal
based on data from the ISOGAL project}\fnmsep
\thanks{Based on observations with ISO, an ESA
project with instruments funded by ESA Member States (especially the
PI countries: France, Germany, the Netherlands and the United Kingdom)
and with the participation of ISAS and NASA}\fnmsep 
\thanks {Partly based on observations collected at the European Southern 
Observatory, La Silla Chile} }

\author{
A. Omont\inst{1}
\and S. Ganesh\inst{1,2}
\and  C. Alard\inst{3,1}
\and  J.A.D.L. Blommaert\inst{4}
\and B. Caillaud \inst{1}
\and E. Copet\inst{5}
\and  P. Fouqu\'e\inst{6}
\and  G. Gilmore\inst{7}
\and  D. Ojha\inst{8,1}
\and  M. Schultheis\inst{1}
\and  G. Simon\inst{3}
\and  X. Bertou\inst{1}
\and  J. Borsenberger\inst{1}
\and  N. Epchtein\inst{9}
\and  I. Glass\inst{10}
\and  F. Guglielmo\inst{1}
\and  M.A.T. Groenewegen\inst{11}
\and  H.J. Habing\inst{12}
\and  S. Kimeswenger\inst{13}
\and  M. Morris\inst{14,1}
\and  S.D. Price\inst{15}
\and  A. Robin\inst{16}
\and  M. Unavane\inst{7}
\and  R. Wyse\inst{17}
}
\authorrunning{A. Omont, et al}
\titlerunning{ISOGAL detection of red giants with low mass loss in the Bulge}

\offprints{A. Omont, omont@iap.fr}

\institute{
Institut d'Astrophysique de Paris, CNRS, 98bis Bd Arago, F-75014 Paris     
\and Physical Research Laboratory, Navarangpura, Ahmedabad 380009, India   
\and DASGAL, Observatoire de Paris, France                                 
\and ISO Data Centre, ESA, Villafranca, Spain                              
\and DESPA, Observatoire de Paris, France                                  
\and ESO, Santiago, Chile                                                  
\and Institute of Astronomy, Cambridge, U.K.                               
\and T.I.F.R., Mumbai, India                                               
\and O.C.A., Nice, France                                                  
\and SAAO, South Africa                  
\and MPA, Garching, Germany                                                
\and Leiden Observatory, Leiden, The Netherlands                           
\and Innsbruck, Austria                                                    
\and UCLA, Los Angeles, CA, USA                                            
\and Air Force Research Laboratory. Hanscom AFB, MA, USA                   
\and Observatoire de Besancon, France                                      
\and The Johns Hopkins University, Baltimore MD, USA                       
}


\voffset 0.3true cm

\date{Received February; accepted May, 1999} 

\maketitle
\sloppy
\begin{abstract}

The ISOGAL project is a survey of the stellar populations, structure,
and recent star formation history of the inner disk and bulge of the
Galaxy. ISOGAL combines 15~$\mu$m and 7~$\mu$m ISOCAM observations
with DENIS IJK$_s$ data to determine the nature of a source and the
interstellar extinction.  In this paper we report an ISOGAL study of a
small field in the inner Galactic Bulge ($\ell$~=~0.0$^{\circ}$,
$b$~=~1.0$^{\circ}$, area~=~0.035 deg$^2$) as a prototype of the
larger area ISOGAL survey of the inner Galaxy. The ISOCAM data are two
orders of magnitude more sensitive than IRAS ones, and its spatial 
resolution is better by one order of magnitude,
allowing nearly
complete and reliable point-source detection down to $\sim$~10~mJy with
the LW3 filter (12-18~$\mu$m) and $\sim$~15~mJy with the LW2 filter
(5.5-8~$\mu$m). More than 90\% of the ISOCAM sources are matched with a
near-infrared source of the DENIS survey.  The five wavelengths of
ISOGAL+DENIS, together with the relatively low and constant extinction
in front of this specific field, allow reliable determination of the
nature of the sources. 

While most sources detected only with the deeper 7 $\mu$m observation are 
probably RGB stars, 
the primary scientific result of this paper is evidence that the most numerous 
class of 
ISOGAL 15~$\mu$m sources are 
Red Giants in the
Galactic bulge and central disk, with luminosities just above or close to 
the RGB tip and
weak mass-loss rates. They form loose sequences in the magnitude-colour
diagrams [15]/$K_s$-[15] and [15]/[7]-[15]. Their large excesses at 15~$\mu$m 
with respect to 2~$\mu$m and 7~$\mu$m is due to circumstellar dust produced by 
mass-loss at low rate ($\dot M_{dust}~\sim10^{-11}$--a few 
$10^{-10}M_{\odot}$/yr). 
These ISOGAL results are the first systematic evidence
and study of dust emission at this early stage (Intermediate AGB and possibly 
RGB-Tip), before the
onset of the large mass-loss phase ($\dot M~\ge~10^{-7}M_{\odot}$/yr). 
It is thus well established that
efficient dust formation is already associated with such low mass-loss
rates during 
this early phase.

About twenty more luminous stars are also detected with larger excess
at 7 and 15~$\mu$m. Repeated ISOGAL observations suggest that the
majority of these are long period variables with large amplitude, probably in
the large mass-loss stage with $\dot M~\ge~10^{-7}M_{\odot}$/yr.


\end{abstract}


\section{Introduction}

The ISOGAL project is a multi-wavelength survey at high spatial
resolution of the inner Galaxy. The general scientific aims are to
quantify the spatial distributions of the various stellar populations
in the inner Galaxy, together with the distribution of the warm 
interstellar medium (ISM). Optical, near-IR and mid-IR (to 15~$\mu$m) data
with near arcsec spatial resolution are being obtained covering the
central Galactic bulge, and sampling the obscured disk within the Solar
circle. Complementary data at other wavelengths are being obtained in
regions of specific interest. To date, most ISOGAL effort has focussed
on a large area broad-band imaging survey at 7~$\mu$m and 15~$\mu$m with
ISOCAM on the ISO satellite, and on complementary DENIS IJK$_s$
observations of the central Galaxy. A full description of the ISOGAL
project will be published elsewhere (Omont et al, in
preparation). First imaging results are available (P\'erault et al 
\cite{Perault96}),
as is a complementary paper to this, discussing late type giants at
somewhat higher Galactic latitudes (Glass et al \cite{Glass99}). In this
paper we discuss first results on the luminous AGB stellar population in the
previously unobserved inner bulge (see also Frogel et al \cite{Frogel99}), from
observations of a field whose line of sight projects some 140pc above the
Galactic centre, on the minor axis.

The AGB stage is one of the most complex phases of stellar evolution,
while at the same time being of crucial importance for nucleosynthesis
and galactic chemical evolution. Although it is clear that mass-loss
dominates the final evolution of AGB stars, the detailed physics of
this process remains rather uncertain. Theoretical progress in the past years 
has
emphasized the relationship between mass-loss and luminosity and hence
thermal pulses, radiation pressure on dust and stellar pulsations of
long period variables (LPV).  The status of recent modelling and the
observational data are fully reviewed by Habing (\cite{Habing96}) [see also 
the proceedings of IAU Symposium 191 edited by
Le Bertre et al. (\cite{LeBertre99})]. Observationally, rates of mass
loss are relatively well documented, especially from millimeter CO
studies, far infrared IRAS results and near infrared studies.  However, this
direct 
knowledge of AGB mass-loss is mostly limited to the solar
neigbourhood. In particular, in the galactic bulge and central disk,
IRAS was able to detect only the relatively few AGB stars with the
largest mass-loss rates $\geq$~10$^{-6}$~M$_{\odot}$/yr. We are indeed
still lacking the observational information to characterize the
influence of metallicity and initial mass on the properties of
mass-loss on the AGB. Near infrared observations, and in particular
the DENIS (Epchtein et al. \cite{Epchtein97}) and 2MASS (Skrutskie et
al. \cite{Skrutskie97}, Cutri \cite{Cutri98}) surveys, can detect
practically all the AGB stars in the Galaxy. However, it is extremely
difficult to both identify an AGB star, and to distinguish between
small to moderate mass-loss rates and patchy interstellar extinction,
solely from near-infrared data in regions of high extinction.  Data at
longer wavelengths, which are more sensitive to the infrared excess
that is a consequence of mass loss, and less sensitive to interstellar
reddening, are required. Mid-infrared post-IRAS space surveys, such as
the present surveys with ISOCAM and MSX (Price et al. \cite{Price97},
Egan et al. \cite{Egan98}) are thus uniquely suited for carrying out
a census of mass-losing AGB stars in the inner Galaxy and for
quantifying the distribution function of mass loss-rates.

Our ISOGAL survey of selected regions of the inner Galaxy, at 15 and 7
$\mu$m (P\'erault et al. \cite{Perault96}, Omont et
al. \cite{Omont99a},\cite{Omont99b}) has a sensitivity 
two orders of magnitude better than IRAS,
and their spatial resolution is better by one order of magnitude.
The main purpose of this paper is to
show that the ISOGAL data, which combine ISOCAM mid-infrared observations with
near-infrared DENIS data, are ideal to identify bulge mass-losing AGB stars,
even with dust mass-loss rates as small as 10$^{-11}$~M$_{\odot}$/yr. 

We analyse here the ISOGAL/DENIS data of a small field (area
0.035 deg$^2$) centered at $\ell$~=~0.0$^{\circ}$, $b$~=~1.0$^{\circ}$ in
the inner bulge. This field is approximately one-half a bulge scale
height down the minor axis, in a previously poorly studied region
mid-way (in $|b|$) between the innermost optical ``windows'' (Glass et al.
\cite{Glass99}) and the Galactic centre. 

The very high stellar density in this field leads to a near
confusion-limited survey, thus providing a maximum number of detected
sources, the majority of them in the bulge and central disk.  The relatively
low and well behaved extinction 
allows reliable identification of mass-losing AGB stars down to the RGB tip 
(K$_0$~$\sim$~8.2 for D~=~8.0 kpc, Tiede et al \cite{Tiede96}). It permits also
easier comparisons with earlier works on (apparently fainter) stars of
the red giant branch (RGB), LPVs and IRAS sources in the more outer bulge (see
e.g. Frogel \& Whitford \cite{Frogel87}, Frogel et
al. \cite{Frogel90}, Tiede et al. \cite{Tiede96}, Glass et
al. \cite{Glass95}, van der Veen \& Habing \cite{Veen90} and references
therein). We can also compare the data with those of the companion paper
(Glass et al. \cite{Glass99}) on ISOGAL observations in two fields of the
Baade Windows. The analysis of the stellar sources is made easier in the
latter by the smaller extinction, the greater distance from the galactic disk
and previous identifications of LPVs from optical surveys.

\section{Observations; Data Reduction and Quality; Cross-Identifications.}
\subsection{ISOGAL Observations}

\begin{figure*} 

\resizebox{\hsize}{!}{\includegraphics{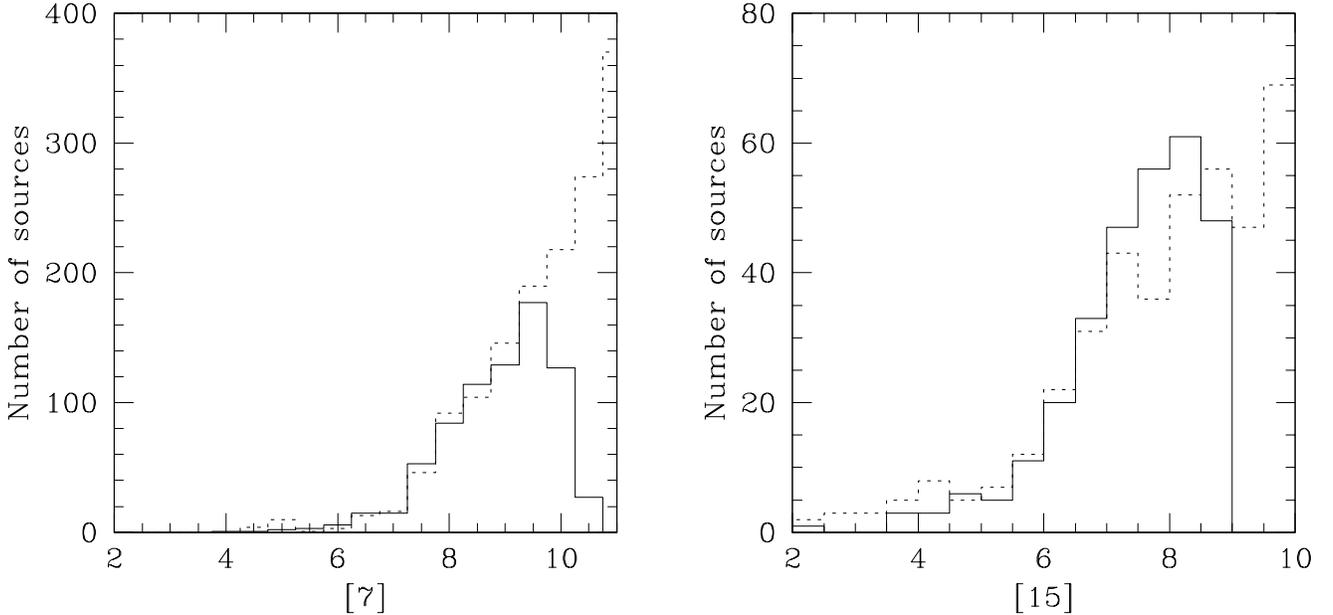}}
\caption{LW2 and LW3 source distributions in half magnitude bins.  Solid lines indicate
the number of detected sources. Dotted lines indicate the approximate expected
number of sources based on LW2$_{exp}$~=~([5$~\times~$K$_s$]~-~10.5)~/~4.0 
and LW3$_{exp}$~=~([1.84$~\times~$LW2]~-~8.5)~/~0.84 for detections in
K$_s$ and LW2 respectively (see Appendix B.4).}
\label{ISO_CMPL}
\end{figure*}

\begin{table}
\caption{Journal of ISOCAM and DENIS observations in the $\ell$~=~0.0$^{\circ}$, $b$~=~1.0$^{\circ}$ field} 

\begin{center}
\begin{tabular}{lcccc}
\hline
Identification&Filter&Pixel Size&Julian Date&Remarks\\
\hline
13600327$^{(b)}$&LW3&$6\arcsec$&2450174&12-18$~\mu$m\\
32500256$^{(b)}$&LW2&$6\arcsec$&2450363&5.5-8.5$~\mu$m\\
83600417$^{(a)}$&LW2&$3\arcsec$&2450873&\\
83600418$^{(b)}$&LW2&$6\arcsec$&2450873&\\
83600522$^{(a)}$&LW3&$3\arcsec$&2450873&\\
83600523$^{(b)}$&LW3&$6\arcsec$&2450873&\\
DENIS 96$^{(a)}$&I,J,K$_s$&$1\arcsec,3\arcsec,3\arcsec$&2450184&\\
DENIS 98$^{(b)}$&I,J,K$_s$&$1\arcsec,3\arcsec,3\arcsec$&2450951&\\
\hline
\label{table1}
\end{tabular}\end{center}
{\it Notes.}$^{(a)}$ Data used in the present paper.\\
$^{(b)}$ These observations have only recently become available and are 
not fully used in the present paper.
\end{table}

This field, $\ell$~=~0.0$^{\circ}$, $b$~=~1.0$^{\circ}$, is one of the 
fields used to quantify the reliability of ISOGAL data. We have thus
at our disposal repeated observations at different dates 
as detailed in Table \ref{table1}. This allows us to check
reliability of detected sources, and additionally to identify (long period) 
variables. Our usual ISOGAL ISOCAM data use $6\arcsec$
pixels. Here we also have observations with $3\arcsec$ pixels (7 \& 15~$\mu$m),
allowing deeper photometry since confusion rather than photon noise
limits the detections. A detailed evaluation will be presented elsewhere in a general assessment of
the quality of the ISOGAL data. We will summarise here a few
conclusions relevant for the present state of data reduction and the
scientific case of this paper. 

The $3\arcsec$ ISOCAM \footnote{see Cesarsky et al. \cite{Cesarsky96} for a
general reference to ISOCAM operation and performances} observations mainly used in this paper 
were performed in revolution 836 (28 February 1998) at 15~$\mu$m
(filter LW3, 12-18~$\mu$m) and at 7~$\mu$m (filter LW2, 5.5-8.5~$\mu$m). 
The two year delay with respect to the IJK$_s$ DENIS
observations should be taken into consideration for the few strongly variable
stars. However, we have another 15~$\mu$m observation at a date
reasonably well matched with that of the DENIS observations.

In addition to the usual problems with the ISOCAM data (glitches, dead
column, time dependant behavior of the detectors), the difficulties of
reduction of the ISOGAL data are more severe for several reasons:
crowding of the fields which is often close to the confusion limit,
highly structured diffuse emission, high density of bright sources which
induce long-lasting pixel-memory effects, integration times per raster
position short compared to detector stabilisation times, etc. Therefore,
a special reduction pipeline was devised (Alard et al. in preparation)
which is more sophisticated than the standard treatment applied to the
ISOCAM data. A detailed discussion of data quality is given in 
Appendix B.


\begin{figure*}
\resizebox{\hsize}{!}{\includegraphics{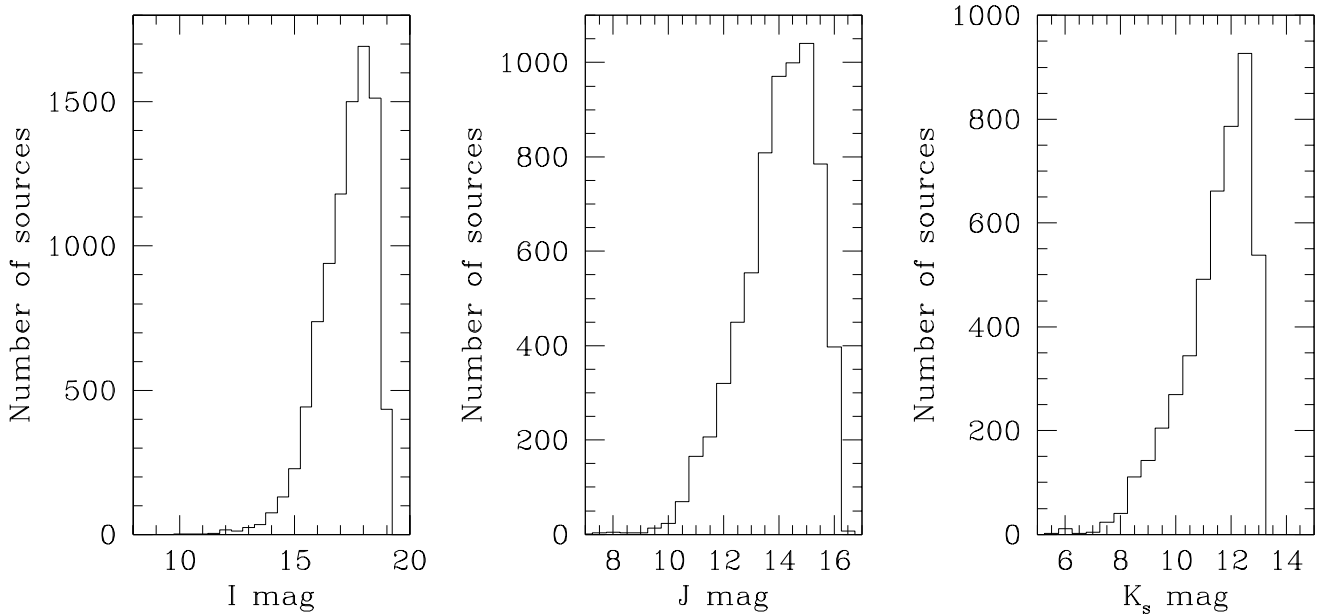}}
\caption{DENIS source counts in I, J, K$_s$ bands in half magnitude bins}
\label{DENIS_HIST}
\end{figure*}

The histograms of the 7 and 15~$\mu$m source counts derived from
the $3\arcsec$ ISOGAL observations are displayed in
Figure~\ref{ISO_CMPL}.  In order to ensure a reasonable level of
reliability, completeness and photometric accuracy, we presently limit
the discussion of ISOGAL data to sources brighter than 8.5 mag (8~mJy)
for LW3 sources and 9.75 mag (11 mJy) for LW2 sources (the fluxes and magnitudes 
used are defined in Appendix A). The source counts in
this field are thus 599 and 282 respectively in LW2 and LW3.

The source density is relatively close 
to the confusion limit for LW2 sources (85 pixels [$3\arcsec \times 3\arcsec$]
per source). However, the LW3 observations are farther from confusion since 
their density is twice smaller than for LW2 sources. 

\subsection{DENIS Observations}

The near infrared data were acquired in the framework of the DENIS
survey, in a dedicated observation of a large bulge field (Simon et
al.  in preparation), simultaneously in the three usual
DENIS bands, K$_s$ (2.15~$\mu$m), J (1.25~$\mu$m) and Gunn-I
(0.8~$\mu$m). Following the general reduction procedures for DENIS
data (Borsenberger et al. in preparation), and after the preliminary
analysis of the same data by Unavane et al. (\cite{Unavane98}), we
optimised the source extraction for crowded fields (Alard et al. in
preparation).  Since for the majority of the ISOGAL sources in this
particular field (giants with little dust and small reddening), the DENIS
sensitivity is much better than that of ISOGAL (by typically 3 magnitudes),
consideration of the faintest DENIS sources is not critical for our purposes.
The histograms of the DENIS K$_s$, J, I sources are shown in Figure
\ref{DENIS_HIST}. The quality of this DENIS data is briefly discussed in
Appendix B. The sensitivity is mostly limited by confusion in the K$_s$ and J
bands. The completeness limit is thus probably close to 11.5 in the K$_s$ band
and 13.5 in the J band (i.e. about two magnitudes lower than in ``normal''
uncrowded DENIS fields). 

One problem with the DENIS data is the saturation of the detectors for
very bright sources. We thus presently do not use the DENIS data for
the 12 sources with K$_s$ $<$ 7, not only for the K$_s$ band, but also
for the I and J bands where the signal is saturated as well. For
slightly fainter sources the corrections for saturation are not yet
optimised and the DENIS photometry will be improved in the future.

The catalog used to make the DENIS astrometry is the PMM catalog,
referenced as the "USNO-A2.0" catalog. The absolute astrometry is then
presently fixed by the accuracy of this catalog (namely $2\arcsec$, with an rms
of $1\arcsec$). The internal accuracy of DENIS observations, derived
from the identifications in the overlaps, is of the order of
$0.5\arcsec$. This excellent astrometry is of course used to improve
that of ISOGAL sources. The DENIS data are also extremely useful to
derive the interstellar extinction toward ISOGAL sources (see Section
3).

\subsection{Cross-identification of ISO and DENIS sources}

\begin{table*}
\caption{Catalog of bright ISOGAL+DENIS sources in the $\ell=0, b=1$ field
($\rm [7] < 7.5$)} 
\begin{center}
\begin{tabular}{rlrrrrrl}
\hline
No.&Name&$I$&$J$&$K_s$&$[7]$&$[15]$&Cross-identifications and comments\\
\hline
   14	&ISOGAL-DENIS-P J174116.3-282957	&     	&10.75	& 8.00	& 
7.23	& 5.86&\\
   23	&ISOGAL-DENIS-P J174117.5-282957	&15.87	& 9.78	& S	&
6.44	& 5.12&\\ 
   61	&ISOGAL-DENIS-P J174122.6-283148	&     	&     	&     	&
4.17	& 2.30& V, IRAS17382-2830\\ 
   94	&ISOGAL-DENIS-P J174126.3-282538	&     	&11.32	& 8.41	&
7.13	& 5.93&\\ 
   99	&ISOGAL-DENIS-P J174126.6-282702	&17.27	&10.41	& 7.37	&
6.00	& 4.47& V\\ 
  108	&ISOGAL-DENIS-P J174127.3-282851	&16.01	&10.03	& 7.43	&
6.31	& 5.01& V\\ 
  119	&ISOGAL-DENIS-P J174127.9-282816	&15.30	& 9.63	& 7.14	&
6.66	& 5.67&\\ 
  124	&ISOGAL-DENIS-P J174128.5-282733	&15.47	& 9.80	& 7.27	&
6.73	& 5.96& V\\ 
  134	&ISOGAL-DENIS-P J174129.4-283113	&14.78	&10.19	& 7.98	&
7.33	& 6.68&\\ 
  148	&ISOGAL-DENIS-P J174130.4-283225	& S    	& 8.14	& S	&
6.57	& 6.67& F\\ 
  158	&ISOGAL-DENIS-P J174131.2-282815	&15.06	& 9.89	& 7.51	&
7.01	& 6.51&\\ 
  212	&ISOGAL-DENIS-P J174134.4-283349	&10.22	& 8.01	& S	&
6.52	& 6.67& F\\ 
  213	&ISOGAL-DENIS-P J174134.4-282922	& S   	& S	& S	&
4.76	& 3.79&\\ 
  218	&ISOGAL-DENIS-P J174134.6-282431	&16.76	&10.53	& 7.96	&
6.51	& 5.57& V\\ 
  220	&ISOGAL-DENIS-P J174134.7-282313	&15.12	&10.61	& 8.34	&
7.49	& 7.73&\\ 
  252	&ISOGAL-DENIS-P J174137.2-282904	&15.82	&10.79	& 8.29	&
7.45	& 6.33&\\ 
  255	&ISOGAL-DENIS-P J174137.4-282630	&15.62	&10.59	& 8.17	&
7.46	& 6.93&\\ 
  270	&ISOGAL-DENIS-P J174138.3-282447	&15.37	&10.35	& 8.11	&
7.46	& 6.74&\\ 
  272	&ISOGAL-DENIS-P J174138.3-282338	&15.35	&10.15	& 7.96	&
7.41	& 6.70&\\ 
  275	&ISOGAL-DENIS-P J174138.6-282743	&16.44	&10.76	& 8.28	&
7.00	& 6.14&\\ 
  289	&ISOGAL-DENIS-P J174139.1-282644	&     	&10.31	& 7.55	&
6.65	& 5.68&\\ 
  296	&ISOGAL-DENIS-P J174139.5-282428	&15.49	& 9.67	& S	&
5.98	& 4.72& V\\ 
  304	&ISOGAL-DENIS-P J174139.9-282520	&15.21	& 9.69	& 7.31	&
6.68	& 4.80&\\ 
  310	&ISOGAL-DENIS-P J174140.1-282220	&11.99	& 9.02	& 7.31	&
7.22	& 7.30& F\\ 
  334	&ISOGAL-DENIS-P J174141.5-282540	&16.46	&10.44	& 8.00	&
7.28	& 6.17&\\ 
  335	&ISOGAL-DENIS-P J174141.5-281930	&14.36	& 9.74	& 7.60	&
7.06	& 7.11& F\\ 
  349	&ISOGAL-DENIS-P J174142.1-283049	&15.68	&11.01	& 8.55	&
7.34	& 6.71&\\ 
  362	&ISOGAL-DENIS-P J174142.6-282641	&17.59	&11.11	& 8.18	&
6.61	& 5.21& V\\ 
  363	&ISOGAL-DENIS-P J174142.7-283116	&     	&11.62	& 8.61	&
6.62	& 5.26&\\ 
  391	&ISOGAL-DENIS-P J174144.5-282034	&     	&10.15	& 8.04	&
7.15	& 6.21&\\ 
  414	&ISOGAL-DENIS-P J174145.4-282824	&15.28	&10.44	& 8.15	&
7.41	& 6.40&\\ 
  433	&ISOGAL-P J174146.5-282259	&     	&     	&     	&
5.08	& 3.83&\\ 
  442	&ISOGAL-DENIS-P J174147.3-282506	&16.21	&10.46	& 8.15	&
7.33	& 6.29&\\ 
  511	&ISOGAL-DENIS-P J174151.4-281739	&16.38	&10.13	& 7.47	&
7.00	& 6.07&\\ 
  522	&ISOGAL-DENIS-P J174151.8-282455	& S   	& S    	& S	&
4.72	& 4.88& F, V\\ 
  529	&ISOGAL-DENIS-P J174152.1-281839	&16.72	&10.57	& 8.07	&
7.44	& 6.65&\\ 
  533	&ISOGAL-DENIS-P J174152.2-281601	&13.58	&10.00	& 7.76	&
6.63	& 6.33& V\\ 
  537	&ISOGAL-DENIS-P J174152.6-282015	&16.30	&10.63	& 8.06	&
7.38	& 6.21&\\ 
  539	&ISOGAL-DENIS-P J174152.8-281720	&14.79	&10.05	& 7.74	&
7.20	& 6.72&\\ 
  548	&ISOGAL-DENIS-P J174153.2-282621	&15.05	& 9.75	& 7.36	&
6.85	& 6.04&\\ 
  550	&ISOGAL-DENIS-P J174153.3-282535	&13.76	& 9.80	& 7.69	&
7.34	& 7.31& F\\ 
  552	&ISOGAL-DENIS-P J174153.4-282027	&13.95	& 9.75	& 7.68	&
7.30	& 7.31& F\\ 
  559	&ISOGAL-DENIS-P J174153.8-282739	&17.23	&11.24	& 8.63	&
7.30	&     &\\ 
  576	&ISOGAL-DENIS-P J174154.6-282133	&16.20	&10.54	& 8.29	&
7.45	& 6.29&\\ 
  579	&ISOGAL-DENIS-P J174154.7-282659	&15.05	& 9.26	& S	&
5.75	& 4.47& V\\ 
  581	&ISOGAL-DENIS-P J174154.8-281731	&16.40	& 9.98	& 7.32	&
6.37	& 5.39& V\\ 
  595	&ISOGAL-DENIS-P J174155.3-281638	&15.66	& 9.59	& S	&
5.99	& 4.49&\\ 
  606	&ISOGAL-DENIS-P J174155.9-282358	&14.12	&10.09	& 8.05	&
7.44	& 6.85&\\ 
  642	&ISOGAL-DENIS-P J174157.5-282237	&15.39	& 9.91	& 7.60	&
6.94	& 5.86&\\ 
  660	&ISOGAL-DENIS-P J174158.7-281849	&     	&10.13	& 7.41	&
5.92	& 4.52&\\ 
  668	&ISOGAL-DENIS-P J174159.3-282554	&11.60	& 8.86	& S	&
6.99	& 6.79& V\\ 
  674	&ISOGAL-DENIS-P J174159.8-281901	&16.80	&10.71	& 8.17	&
7.49	& 6.45&\\ 
  682	&ISOGAL-DENIS-P J174200.3-282303	&12.20	& 8.16	& S	&
5.67	& 4.93& F?, V\\ 
  701	&ISOGAL-DENIS-P J174201.9-281802	&16.76	&10.58	& 8.01	&
7.37	& 6.35&\\ 
  716	&ISOGAL-DENIS-P J174202.8-282124	&16.16	&10.39	& 8.10	&
6.41	& 5.75& V, Terzan V 3126\\ 
  723	&ISOGAL-DENIS-P J174203.2-282107	&12.09	& 8.36	& S	&
6.43	& 6.32& F\\ 
  725	&ISOGAL-DENIS-P J174203.7-281729	&16.98	&10.52	& 7.61	&
5.89	& 4.58& V\\ 
  732	&ISOGAL-DENIS-P J174204.3-282137	&13.52	& 9.33	& 7.33	&
6.79	& 6.92& F\\ 
  753	&ISOGAL-DENIS-P J174206.8-281832	&17.59	&10.68	& 7.65	&
5.49	& 3.87& V\\ 
  791	&ISOGAL-DENIS-P J174213.8-281827	&12.85	& 8.65	& S	&
6.11	& 5.63& F\\ 
  794	&ISOGAL-DENIS-P J174215.1-281850	&16.34	&10.37	& 7.78	&
6.86	& 5.85& \\ 
  799	&ISOGAL-DENIS-P J174216.3-281947	&15.49	&10.34	& 8.05	&
7.18	& 6.71&\\ 
\hline
\label{table2}
\end{tabular}
\end{center}
\vspace{-0.6cm}
{{{\it Note.} F = Foreground or suspected foreground star, V = Variable or suspected variable star, S = Saturated source}}\\

\end{table*}

\begin{figure}
\resizebox{\hsize}{!}{\includegraphics{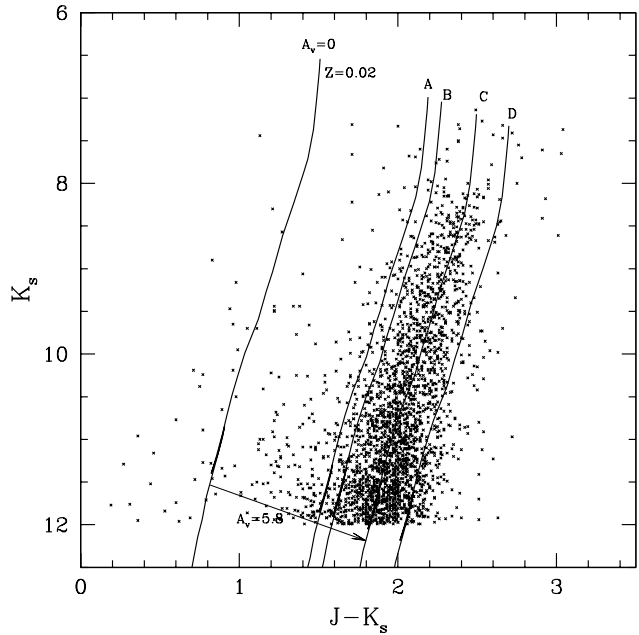}}
\caption{Colour magnitude diagram (J-K$_s$) / K$_s$ for all unsaturated DENIS
sources in the field. An isochrone (Bertelli et
al. \protect\cite{Bertelli94}), placed at 8\,kpc distance, is shown
for a 10\,Gyr population with Z=0.02. The near-infrared colours of
this isochrone have been computed with an empirical $\rm
T_{eff}-(J-K)_{0}$ colour relation built by making a fit through
measurments 
see Ng et al (in preparation)
for details about this relation. The labels
A,B,C,D identify the isochrones shifted by A$_{\rm V}$ of 4, 4.5, 5.8
and 7 respectively.}
\label{K_JK_DEN}
\end{figure}

\begin{figure}
\resizebox{\hsize}{!}{\includegraphics{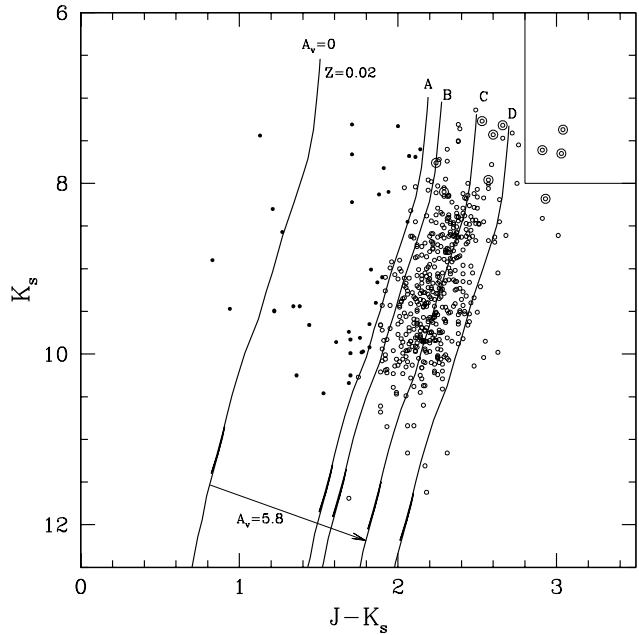}}
\caption{Colour magnitude diagram (J-K$_s$)/K$_s$ for unsaturated DENIS
sources with [7] or [15] counterparts in the field.  Filled circles represent
the foreground sources with consistent data in the other diagrams.
 Suspected variables (see text and Section 6) are
indicated additionally by large open circles.  The labels A,B,C,D and isochrones
are as described in Figure $\ref{K_JK_DEN}$.  The region where J-K$_s$ is 
indicative of high mass-loss 
AGB (see Section 6)
is delimited by the boxed region to upper right.}
\label{K_JK_ISO}
\end{figure}

Cross-identifications of LW3 and LW2 sources, between themselves and
with DENIS sources, provide the multi-colour data which allow
discussion of the nature and properties of individual sources which
makes up the bulk of this paper (Figures \ref{K_JK_DEN} to
\ref{7_K7} discussed below).  The cross-identification process is
also useful for determining data quality. We have now routine standard
procedures for ISOGAL-ISOGAL and DENIS-ISOGAL cross-identifications (see 
Appendix B). A substantial fraction of the ISO sources have thus been
identified with DENIS sources (93\% and 84\% for LW2 and LW3 sources 
respectively).

Cross-identifications are essential to establish the reliability of
the ISOGAL detections. Indeed, because of possible residual artifacts,
mainly due to pixel memory or of noise peaks simulating sources, we
consider that the reality of a weak ISOGAL source is not yet well
warranted here if it is not confirmed by another detection, either in
the other ISOGAL band, or in the K$_s$ band. Only 9\% of LW3 sources
are not associated with an LW2 or a K$_s$ source.
The proportion of unassociated LW2 sources is slightly smaller, 5\%.

In order to check the completeness of LW3 sources, we can use the more
sensitive LW2 observations; DENIS K$_s$ detections can be used in a similar
way to estimate the completeness of LW2 sources (see Appendix B and 
Figure~\ref{ISO_CMPL}). The completeness is probably close to 80\% at least, 
for $[7]~<~9.5$ and for $[15]~<~8.5$.
A more detailed analysis of the source surface density, and its
implications for the structure of the Galactic bulge, will appear
elsewhere, following more sophisticated modelling of source
incompleteness as a function of position in this and other fields.
For present purposes such an incompleteness is not a limiting factor.

The quality of ISOGAL photometry has been checked in this field and
others by repeated observations (see Appendix B and Ganesh et al. in 
preparation). The uncertainty thus proved to be better than $\sim 0.2$ mag rms
above $\sim 15$ mJy in both bands. However, there is not yet a good standard
procedure to fully correct for the detector time behaviour effects for fields
such as ISOGAL ones with strong sources and background. Because of that and of
confusion, our photometry is thus still uncertain by a few tenths of a
magnitude systematically. 

In conclusion, we consider the reliability of the existence of most of the
sources discussed to be well established. The completeness is also 
well characterised. However, the photometric accuracy can still be
improved.

Table \ref{table2} gives a catalogue of bright ISOGAL sources ($[7]~<~7.5$), with three-band
DENIS associations and identification of foreground sources and of candidate
variable stars. A complete catalogue of all ISOGAL sources will be available
at CDS by October 1999, when the data reduction is improved. 

%

\section{Near infrared data and interstellar extinction}

The data at five wavelengths available for most of the sources allow
in most cases a good characterisation of the ISOGAL sources, with some
redundancy, as well as of their interstellar reddening. The very large
stellar density in the inner bulge and central disk brings a considerable
simplification by ensuring that the majority of the sources are located within 
it. 
In addition, it happens that the interstellar extinction is
relatively small on this line of sight and nearly constant for this
whole small field. The discussion of the nature of the sources and of
their properties, such as mass-loss, is thus much easier.

The multi-dimensional analysis of magnitude-colour space defined by
these data allows one to visualise and to determine the source
properties. Deferring detailed discussions to the next sections, we
limit this section to a general presentation of these diagrams and of
the general information they provide about interstellar extinction,
circumstellar dust emission and circumstellar absorption.

The K$_s$/J-K$_s$ magnitude-colour diagram of all DENIS sources in our
field (Figure \ref{K_JK_DEN}) shows a remarkably well-defined bulge
red giant sequence shifted by fairly uniform extinction of A$_{\rm
V}$=5.8$\pm$1 mag. with respect to the reference
K$_s$$_o~vs~{(J-K_s)}_o$ of Bertelli et al. (\cite{Bertelli94}) 
\footnote{The isochrones by Bertelli et al. have been computed in the ESO 
system using
the ESO filter curves. From NIR spectra for a sample of oxygen-rich M
stars and carbon stars, K-K$_s$ values have been computed. The differences
are very small (on average about 0.04 mag for the M giants), so 
concerning the internal dispersion in K we can neglect it.
This result is in agreement with Persson et al. (\cite{Persson98})
who presented a new grid of infrared standard stars in J, H, K and K$_s$.}
%
%
with Z~=~0.02 and a distance modulus of 14.5 (distance to Galactic Centre
8~kpc; we have assumed that [A$_{J}$-A$_{K_s}$]/A$_{\rm V}$~=~0.167). 
Most of the extinction should thus be associated with
interstellar matter outside of the bulge. We have checked from the
individual values of A$_{\rm V}$ and the DENIS source counts that
there is apparently no strong spatial variation of the
extinction in this field. The ISOGAL sources with anomalously low values of A$_{\rm
V}$ are probably foreground. They are visible in Figure
\ref{K_JK_ISO}, which shows the subset of the K$_s$/J-K$_s$ sources of
Figure \ref{K_JK_DEN} which were also detected at longer wavelengths.
There is of course some uncertainty for the marginal cases: for bright
sources with good photometry (K$_s$~$<$~10), those located left of
line A (A$_{v}~<~4$) are almost certainly foreground, while those to
the right of line B (A$_{v}~>~4.5$) are very probably in the ``bulge'',
and the case is uncertain for those between lines A and B. The case of
foreground ISOGAL sources will be further discussed below. Those with
A$_{v}~<~4$ and with consistent data in the other diagrams are identified by
special symbols in the various diagrams.

The distribution of sources about line C is dominated by photometric
errors, and residual extinction variations. There is no evidence for a
significant background population, more highly reddened, in the disk
beyond the Galactic centre, though a few such sources may be present.
Similarly, it is difficult from just these data to identify any
intrinsic red J-K$_s$ excess in sources below the RGB-tip, which is near
K$_{s,0}~=~8.2$ (Tiede et al \cite{Tiede96}).  The J-K$_s$ excess of the six
bright sources much 
redder than line D is very probably related to an intrinsic J-K$_s$
excess generated by a relatively thick dusty circumstellar shell, as
confirmed by the very large value of K$_s$-[15] for these sources (see
Section 6). A few other sources, just redder than line D in Figure
\ref{K_JK_ISO}, might also have an intrinsic J-K$_s$ excess.  Only
in the situation of relatively small, foreground, and uniform
interstellar extinction, could even a fraction of the AGB stars with high
mass-loss be identified, and their mass-loss characterised, from the
near-infrared DENIS (or 2MASS) data alone. In general, longer
wavelength data are critical.

The I band data, when they exist, can provide additional interesting
constraints. 
However, the much
larger spread of intrinsic I-J values in the bulk of the distribution,
compared to J-K$_s$ complicates the identification of foreground sources from
I-J data alone. 
While (J-K$_s$)$_{\rm o}$ is confined to a very small range
($\sim~$0.5 mag) for most sources, it is well known (see, e.g., Frogel
\& Whitford \cite{Frogel87}, Appendix A) that there is a large spread
in the I magnitudes of bulge AGB giants and hence in (I-J)$_{\rm
o}$, that we find ranging over more than 2 magnitudes. 
Such a spread is certainly related to the behaviour of the TiO
absorption bands, and hence probably to the metallicity. However,
there is as yet no very detailed modelling of this behaviour.

The average value of I-J
increases by about 1.5 mag. along the intermediate-AGB sequence defined in 
Section 5.

\section{The Nature of the ISOGAL Sources}

As discussed below, 
the most numerous classes of sources detected both at 7 \& 15~$\mu$m in the
ISOGAL survey are probably ``bulge'' intermediate AGB stars or RGB tip stars 
with low mass-loss, and high mass-loss rate
AGB stars
($\dot M~\ge~10^{-7}M_{\odot}$/yr). These are discussed in detail below. In
this section we consider first minor populations in the survey. 

There are practically no good young star candidates among ISOGAL sources in 
this field. They should be found among sources with 
large 15$\mu$m excess that are too faint to be AGB stars with large mass-loss. 
There are no really convincing cases in the diagrams of Figures \ref{15_K15} \& \ref{15_715} (
however, 
see Section 5). This is 
consistent with the relatively small value of A$_v$, indicating that there is 
no very thick molecular cloud on the line of sight.

\subsection{Foreground stars with little or no reddening}

\begin{figure}
\resizebox{\hsize}{!}{\includegraphics{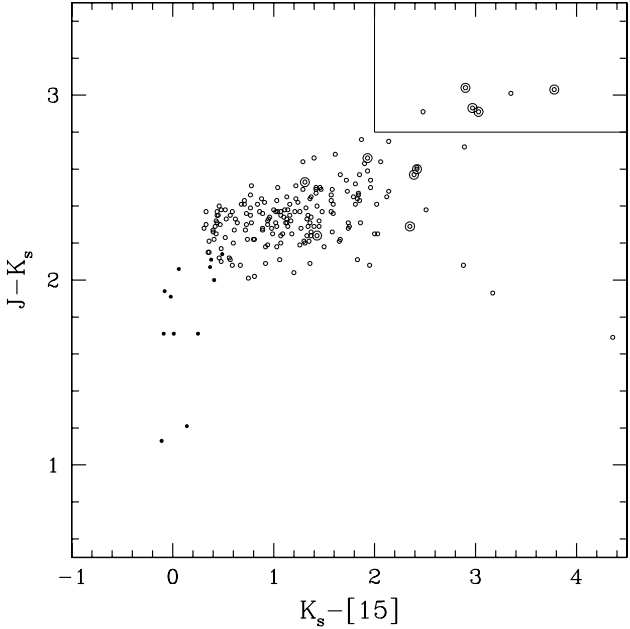}}
\caption{J-K$_s$/K$_s$-[15] colour-colour diagram of LW3 sources with
unsaturated DENIS counterparts. All symbols are as in Figure
$\ref{K_JK_ISO}$.  
}
\label{JK_K15}
\end{figure}

Some 40
stars ($\sim7\%$ of ISOGAL sources with DENIS counterparts) which
lie to the left of line A (A$_{\rm V}$$\sim$4) in Figure
\ref{K_JK_ISO} are probably foreground stars, in front of the main
line of sight extinction. This is independently confirmed by another
colour for most of them. The brightest 19 stars, with
K$_s$$ \la$ 9, are detected in LW3 with colours 0.1$ < $K$_s$-[15]$ < $0.5 
and apparent magnitudes consistent with their
being foreground disk giants. The fainter sources are
consistent with being either M giants with very low reddening, or
moderately reddened disk K giants.  

One should add to these probable foreground stars, a few very bright
sources saturated in the DENIS data. Seven such stars are thus
tentatively identified in the ISO data. Let us stress that the identification 
of foreground stars is more difficult for bright sources (K~$\la$~8) because 
the intrinsic colour (J-K$_s$)$_0$ is more uncertain. 
Altogether, we have thus
identified about 8\% of the ISOGAL sources as foreground stars. They
are distinguished by special symbols in Figures \ref{K_JK_ISO} to
\ref{7_K7}.

\subsection{7$\mu$m sources without 15$\mu$m detections}

As described above, the ISOGAL sensitivity is much greater in the LW2
band than in the LW3 band for red giants with no or little dust. The
number of sources detected in LW2 is more than twice that in LW3. Most
sources detected with LW2 and not with LW3 are fainter in the LW2 and
K$_s$ bands than are the detected LW3 sources (see Figure \ref{7_K7}).

We are able to define the nature of these sources reliably, since the
interstellar extinction is well characterised on this line of
sight. Few, if any, of the sources detected only at 7~$\mu$m can have intrinsic
infrared excess, from circumstellar dust, or they would have been
readily detected at 15~$\mu$m. The foreground sources are identifiable 
from combination of the DENIS and 7~$\mu$m flux. 
Thus, one may isolate 
those sources which are predominately bulge RGB sources, below the RGB
tip (see Figure \ref{7_K7}). An analysis of the bulge density distribution of
both AGB and RGB 
stars, based on their surface density distribution, will be provided
elsewhere. This analysis however requires very careful modelling, as
the faint source counts are strongly affected by incompleteness (see e.g. 
Unavane et al \cite{Unavane98}).

\section{Mid-Infrared Data and Intermediate AGB stars}

The main additional information in the ISO mid-infrared bands with
respect to the shorter wavelength data solely from DENIS, 
is the much increased sensitivity to emission from cold
circumstellar dust. This is well exemplified by the J-K$_s$/K$_s$-[15]
colour-colour diagram of Figure \ref{JK_K15}. While the range of
J-K$_s$ values is restricted to $\sim$0.5~mag for most sources (with
another 0.5~mag for a few sources), K$_s$-[15] ranges from 0 to 2.2
for the bulk of the sources (with an extension up to 4 magnitudes for
a few sources). The colours [7]-[15] and K$_s$-[7] (Figures
\ref{15_715} \& \ref{7_K7}) also display large ranges of excess,
although somewhat smaller than for K$_s$-[15].

As we discuss now, only the presence of circumstellar dust can explain
such a large excess; only a portion of it can be attributed to the
very cold photosphere.

\begin{figure}
\resizebox{\hsize}{!}{\includegraphics{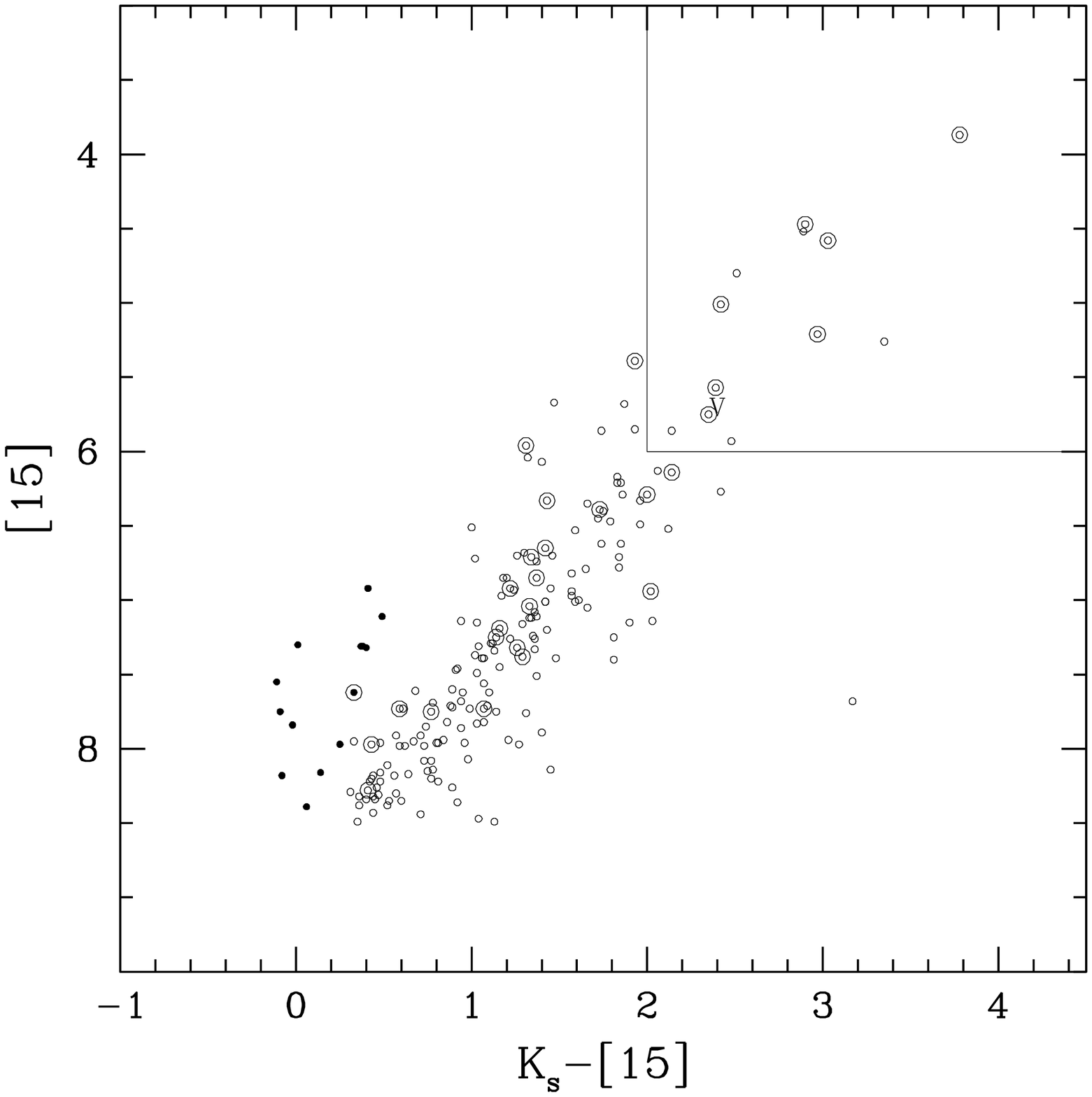}}
\caption{[15]/K$_s$-[15] magnitude-colour diagram of ISOGAL sources
with unsaturated DENIS counterparts. Symbols are as in Figure
$\ref{K_JK_ISO}$.  
All the possible candidate variables selected by the method described in 
Section 6 are indicated by large open circles. However, those with [7]~<~7.0 
are considered as dubious and they are discarded in the other figures. 
The Terzan variable (see text) is refered by the
symbol V in this figure.
The region of high mass loss AGB is demarcated by
the boxed region.}
\label{15_K15}
\end{figure}

\begin{figure}
\resizebox{\hsize}{!}{\includegraphics{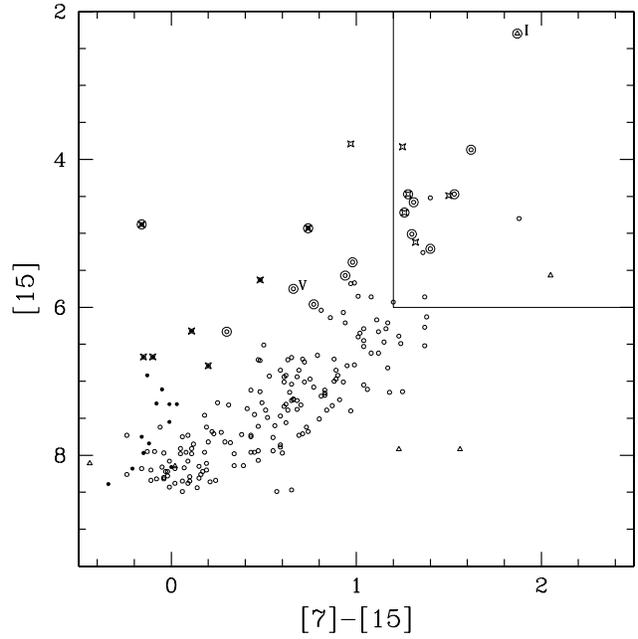}}
\caption{[15]/[7]-[15] magnitude-colour diagram of ISOGAL
sources. Sources saturated in DENIS observations are shown by filled
(foreground) and open (bulk) crosses in this figure. Sources without
DENIS counterparts are shown by open triangles. The Terzan variable is
denoted by V and the IRAS source by I (see text).  The region of high
mass loss AGB is demarcated by the boxed region. All other symbols are
as in Figure $\ref{K_JK_ISO}$. }
\label{15_715}
\end{figure}

\begin{table*}
\label{Table3}
\caption{ Values of colours and magnitudes for the base 
and the tip of the mass-loss AGB sequence
(``intermediate-AGB sequence'') in the magnitude-colour diagrams
Figures \ref{15_K15},
\ref{15_715} \& \ref{K_K15}.}

\begin{center}
\begin{tabular}{lccccccccc}
\multicolumn{10}{c}{\ }\\
\hline
\multicolumn{10}{c}{\ }\\
\multicolumn{1}{c}{} & 
\multicolumn{1}{c}{K$_s$-[15]} &
\multicolumn{1}{c}{[7]-[15]} &
\multicolumn{1}{c}{[15]} &
\multicolumn{1}{c}{K$_s$} &
\multicolumn{1}{c}{(K$_s$-[15])$_0$} &
\multicolumn{1}{c}{K$_{s0}$} &
\multicolumn{1}{c}{M$_K$$^a$} &
\multicolumn{1}{c}{M$_{bol}^{(b)}$} &
\multicolumn{1}{c}{L(L$_{\odot}$)}\\
\\
\hline
\multicolumn{3}{c}{\ }\\
Tip  &  2.2 & 1.4 & 5.8 & 8.0 & 1.8 & 7.5 & -7.0 &  -4.0 &  3200\\
Base & 0.4 &   0.0 &  8.5 &   8.9  &   0  &  8.4 & -6.1 & -3.1 & 1400\\  
\multicolumn{10}{c}{\ }\\
\hline
\end{tabular}
\end{center}

\noindent
{\it Notes.}
$^{(a)}$ with a distance modulus of 14.5 (D~=~8~kpc)\\
$^{(b)}$ with the K bolometric correction M$_{bol}$~-~M$_{\rm
Ks}$~=~3.0 (Groenewegen  \protect\cite{Groenewegen97}), which yields
M$_{bol}$$\sim$K$_s$-12 in this field.

\end{table*}

\begin{figure}
\resizebox{\hsize}{!}{\includegraphics{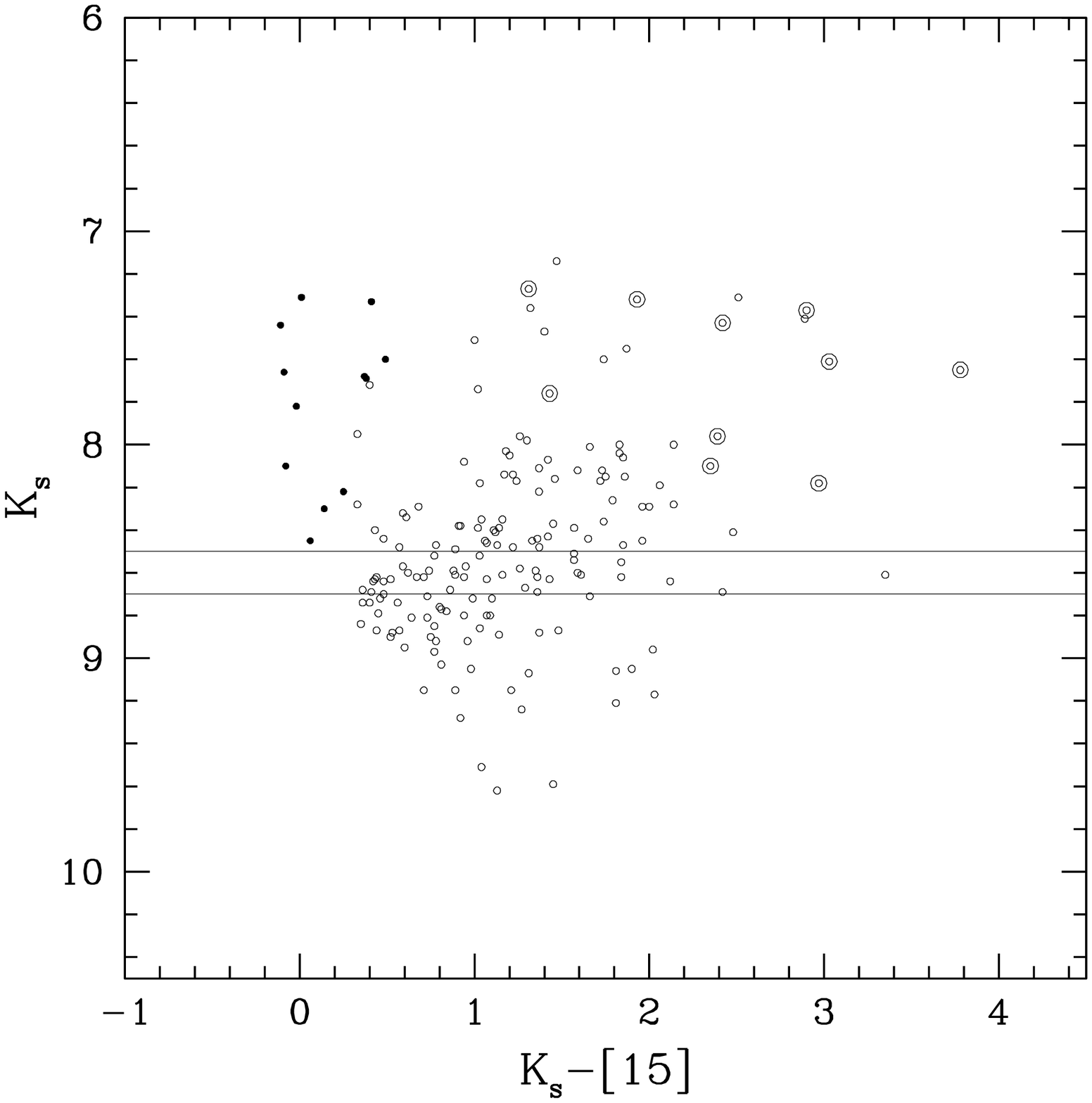}}
\caption{K$_s$/K$_s$-[15] magnitude-colour diagram of ISOGAL sources 
detected both in LW2 and LW3, 
with unsaturated DENIS counterparts. The 
approximate position of the RGB tip (taking into account the
interstellar extinction in this field) is shown by the solid lines at 
K$_s$=8.7 (K$_0$~$\sim$~8.2, Tiede et al. \cite{Tiede96}, A$_K$~$\sim$~0.5) and 
K$_s$=8.5 (K$_0$~$\sim$~8.0, Frogel et al. \cite{Frogel99}). 
All symbols are as in Figure
$\ref{K_JK_ISO}$.}
\label{K_K15}
\end{figure}

In the magnitude-colour diagrams [15]/K$_s$-[15] and [15]/[7]-[15] 
(Figures \ref{15_K15} \& \ref{15_715}, 
respectively), the 
majority of the sources follow a loose linear sequence.
Characteristic values of the colours and magnitudes corresponding to the two
ends of this 
sequence are given in Table~3. The magnitude of the lower end of the
sequence accidently coincides with the ISOGAL sensitivity at 15$\mu$m. It 
is almost exactly that of the tip of the bulge RGB (~$K_{\rm
o} \sim$~8.2, Tiede et al \cite{Tiede96},~$K_{\rm
o} \sim$~8.0, Frogel et al \cite{Frogel99})
\footnote{The magnitude spread from the line 
of sight effect of the bulge is 
very model-dependant. Perhaps the most reliable first estimate can be derived
from recent models of the COBE photometry (Binney et al. \cite{Binney97}),
which Glass et al (\cite{Glass99}) show are reasonably consistent with
the ISOGAL source counts in the inner Galaxy. These analyses derive a bulge
density profile which is approximately exponential with scale length smaller
than 300pc. For an adopted galacto-centric distance of 8kpc, this introduces a
width of $\sim$ 0.1mag per scale length. 
In their study of Sgr I Miras, Glass et
al 1995 have found a dispersion sigma of 0.35 mag from
the regression line in the K logP diagram. This gives an upper limit for
the dispersion caused by front-to-back spread, at least in the Sgr I region.
Thus, line of sight spreads in the
photometry, for fields near the minor axis where systematic bar-induced
effects are unimportant, are probably small compared to other uncertainties.
Of course, the magnitude spread is much larger for the minor, but non 
negligible, number of sources of the central disk: $\sim$0.9 mag per 2.7 kpc 
scale length}.

Since there is some uncertainty about the position of the base of the 
sequence with respect to the RGB tip, 
it is difficult to know whether the 15$\mu$m sources with the smaller infrared 
excess (K$_s$-[15]$~\la~1$) are intermediate-AGB or RGB-tip stars. However
, 
most of the sequence with larger 15
$\mu$m excess (K$_s$-[15]$~\ga 1$) 
seems to correspond to AGB stars up to $\sim$1~mag in K$_s$
brighter than the 
RGB tip. This K$_{s0}$ magnitude range, 7.5~-~8.5, corresponds to
M spectral types from M6 to M9 (see Table~3A of Frogel \& Whitford
\cite{Frogel87} and Figs 10 and 12 of Glass et
al. \cite{Glass99}). Since these stars are fainter by one or two
magnitudes in K$_s$$_{\rm o}$ or M$_{bol}$ than the few very luminous AGB stars
discussed in Section 6, we propose to describe this sequence as
``the intermediate-AGB mass-loss sequence''.

There are also some stars with large 15$\mu$m excess (K$_s$-[15]$~>~1$) 
apparently significantly below the RGB tip (see Figure \ref{K_K15}). However,
we have checked that the majority (8 out of 13) have very poor photometry 
because of blends. The nature of the few remaining cases is unclear: 
photometry or association problems, background AGBs, young stars or red giants 
below the RGB tip (AGB or RGB) with mass-loss?


In order to explore the amount of circumstellar dust involved and its
properties, we have used the models developed by one of us (MG) which
calculate absolute magnitudes within the relevant ISOCAM and DENIS
filters. 
The most robust conclusion from the models is confirmation of
the need for circumstellar dust to achieve such large infrared
excess with respect to photospheric emission. Without dust the K$_s$-[15]
colours for giant spectral types M5, M8 and M10 are only 0.15, 0.55 and 1.07,
respectively.  As 
concerns the specific dust model, in the absence of detailed
information, the simplest assumption is a time-independent dust
mass-loss rate $\dot M_{dust}$. Of course the value of $\dot M_{dust}$
inferred from the DENIS-ISOGAL colours strongly depends on the assumed
intrinsic dust properties. 
Depending on these properties, the dust mass-loss rate of the tip of the 
intermediate AGB sequence ranges from 
$\dot M_{dust}$ = $\sim$10$^{-10}~M_{\odot}$/yr to 
$\sim$5 10$^{-10}~M_{\odot}$/yr.
The infrared colours of the beginning of the
intermediate AGB sequence imply mass-loss rates 10-30 times smaller
than for the tip.

An appropriate value of the dust-to-gas ratio during mass loss remains
an open question. The range of spectral type and of mass-loss rate
discussed above is typically the domain of validity of the Reimers
formula (1975) for the total mass-loss rate $\dot M$. For M$_{bol}$ =
$-$4, this formula gives $\dot M \sim 6~10^{-8}$M$_{\odot}$/yr. If one
assumes that this result, derived in the solar neighbourhood, can
still be applied to bulge stars with the same luminosity, it yields a
gas-to-dust ratio in the range $\sim$100-500, depending on the dust
properties. However, the gas-to-dust ratio should be significantly smaller at 
the base than at the tip of the ``intermediate-AGB sequence'' of 
Figure \ref{15_K15}.

The nature of the dust of bulge 
stars with weak mass-loss will be discussed in a 
forthcoming paper (Blommaert et al. in preparation) from the 
ISOCAM-CVF spectral observation (5-16.5~$\mu$m) of a few 
$3\arcmin \times 3\arcmin$ ISOCAM fields in the bulge.

Let us stress that the intermediate AGB
stars in the central Galactic bulge with
low mass-loss rates (several 10$^{-9}$ to close to 10$^{-7}$~
M$_{\odot}$/yr) seem rather similar in their properties to Solar
Neighbourhood IRAS AGB stars with typical [12]-[25] colours in the
range 0.2-1 (Guglielmo 1993, unpublished PhD thesis, Hacking et al. \cite{
Hacking85}).

\section{Luminous Bulge AGB Stars}

The apparently brightest stars detected by ISO at mid-infrared
wavelengths form a loose group of $\sim$~30 sources (see Figure \ref{15_715}) 
taking into account those 
saturated in DENIS bands,  in the
various magnitude-colour diagrams. They are brighter than 
the tip of the sequence of ``intermediate-AGB'' sources, which we have
shown are AGB stars  with low  mass-loss rates.
These stars have  [15]$ \la$ 6, [7]$ \la$ 7, and  K$_s$$ \la$ 8
in the various figures. Note that with conversion between K$_s$
and M$_{bol}$ (Table~3), this limit corresponds to M$_{bol}
$$ \la$ -4.

    A large proportion ($\sim$$2/3$) have at least two very red
colours among K$_s$-[15]$ > $2.0, [7]-[15]$ > $1.2, K$_s$-[7]$ > $1.0 and
J-K$_s$ $ > $2.8, characteristic of larger mass-loss than for the
intermediate-AGB sequence. It is tempting to identify them with the
onset of the AGB ``large mass loss'' at $\dot
M$$ \sim $10$^{-7}~M_{\odot}$/yr. It is known that such a strong wind
is classically associated with long period variability (LPV, see,
e.g. Habing \cite{Habing96} and references therein). From the SIMBAD
data base, we have found two LPV stars previously identified in this field,
\object{IRAS17382-2830} and \object{Terzan V3126} (Terzan \& Gosset
\cite{Terzan91}). IRAS17382-2830 is an OH/IR star with S$_{25\mu
m}$/S$_{12\mu m}$~=~2.12. It is denoted by ``I'' in Figure \ref{15_715}.
Remarkably, this source (together with two other very bright 15$\mu$m
sources) is not detected in our 1996 DENIS observations. Its derived colours
are extremely red: K$_s$-[7]$ > $7, K$_s$-[15]$ > $9.  Such very red colours 
are confirmed by the 1998 DENIS observations 
(see Table 1) where the source is detected, 
giving K$_s$ = 11.7, K$_s$-[7]$ = $7.56 and K$_s$-[15]$ = $9.43 
(Schultheis et al, in preparation). These extreme near-IR colours are
consistent with its very cold 12/25$\mu$m IRAS colour (see, e.g.,
Blommaert et al. \cite{JBlommaert98}).  

\begin{figure}
\resizebox{\hsize}{!}{\includegraphics{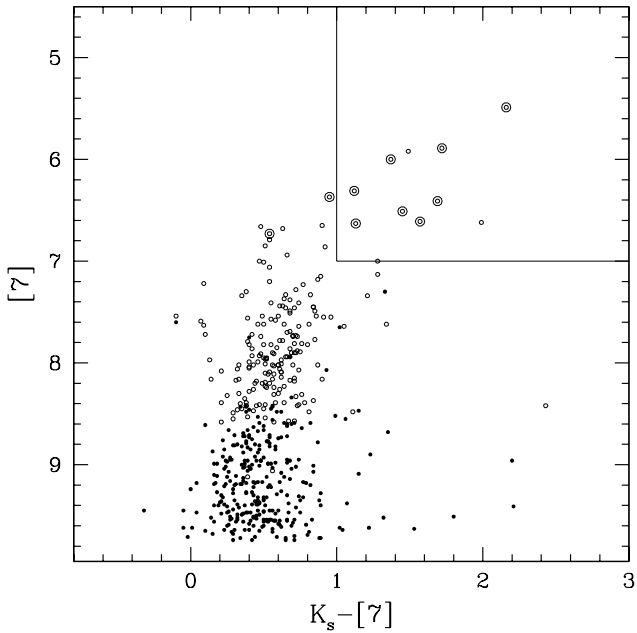}} 
\caption{[7]/K$_s$-[7] magnitude-colour diagram of all LW3 sources
with unsaturated DENIS counterparts. Filled circles represent sources
without a detection at LW3. Suspected variables are indicated additionally by
large open circles.
The region of high
mass loss AGB is demarcated by the boxed region.}
\label{7_K7}
\end{figure}

In order to investigate variability in this field we compared the
observations performed with $6\arcsec$ pixels, at two different dates
(see Table 1) with both LW2 and LW3 filters. 
We consider that a bright source is a
candidate to be considered for variability, when 
there is a 3$
\sigma$ difference in one band, or consistent weaker indications in both
bands. The sources selected in this way are displayed with special
symbols in Figures \ref{K_JK_ISO} to \ref{7_K7}.  We emphasise
that this is just a positive indication in favor of variability, but
without any rigorous statistical meaning. In particular, the significance of 
the candidate variables on the intermediate AGB sequence (Figure \ref{15_K15}) 
is unclear, since they are not confirmed by variability in DENIS data (Schultheis 
et al. in preparation), and it is known that there is no strong
variables on this sequence in Baade's Window (Glass et al \cite{Glass99}). 
On the other hand, with
observations at only three epochs, we can miss a few variables. This is the
case for example for the known variable Terzan V 3126. 

A first striking feature in the distribution of these suspected
variables is their high proportion in the regions of the
magnitude-colour diagrams corresponding to the high mass-loss AGB
stars defined above ($\dot M$$ \ga $10$^{-7}M_{\odot}$/yr)
and delimited in Figures \ref{K_JK_ISO}, \ref{JK_K15},
\ref{15_K15}, \ref{15_715}  and \ref{7_K7}. In all cases these
candidate variables are at least 50$\%$ of the stars found there.
Their proportion exceeds 80$\%$ in Figure \ref{7_K7} ([7]$ < $7.0, 
[7]-[15]$~>~$1.0. This region of the [7] $vs$
[7]$~-~$[15] plane has been shown from analysis of ISOGAL data by Glass
et al. (\cite{Glass99}) to be best correlated with the LPV phenomenon
in Baade Window fields. This identification of luminous variable stars using 
ISO photometry has been confirmed 
by preliminary comparisons of DENIS data at two epochs
(Schultheis et al, in preparation), in particular for most individual stars of 
this field. Altogether, we have 16 candidate LPVs
among the 33 brightest stars ([7] $ < $ 7.0).

Most of the sources with [7] $ < $ 7 
which are not candidate LPVs are grouped in the region
6.4 $ < $ [7] $ < $ 7, 0.4 $ < $ K$_s$ - [7]  $ < $ 1. Out of 12 sources there, 
only two are candidate variables. Of course, we have not enough data to claim 
with any certainty that any given star among the 10 other is not a variable. 
However, it is clear, from comparison with the similar group with 
K$_s$ - [7] $ > $ 1 where almost all stars are candidate variables, that the 
majority of these 10 stars are not strong variables. The relationship of this
apparently non-variable group with AGB stars of similar K magnitude
which are LPVs, and the relationship with less bright AGBs with
however similar colours is still unclear (see also Glass et al \cite{Glass99}). 
It will be interesting to
check in particular whether the difference between variable and
non variable AGB stars in apparently similar evolutionary states is
related to differences in metallicity or initial mass, or to a
difference in mass loss history on the AGB. Another possibility is that most 
of the ``non variable'' bright stars are in the inner disk with $D~<~8~kpc$, 
since their values of J-K$_s$ are smaller than the average value (curve C in 
Figure \ref{K_JK_ISO}) for most of them.

\section{Conclusion}

ISOGAL, combining ISOCAM and DENIS data, is providing the first
detailed and systematic mid-infrared study of an inner bulge field
with sufficient sensitivity to isolate the entire AGB population.  The
ISOGAL performance allows a breakthrough in the analysis of infrared
stellar populations, with an improvement of two orders of magnitude
with respect to IRAS. The density of detected sources is close to the
confusion limit with 3$\arcsec$ pixels at 7~$\mu$m. As expected, most
of the sources are M giants of the bulge or central disk. In the low reddening
fields 
analysed to date practically all bulge giants have near-infrared DENIS
counterparts.  This high proportion of associations is possible
only because of the very good DENIS astrometry and the relatively good
astrometry with ISOCAM.  Sources detected above our
completeness limit at 15~$\mu$m are mainly just above the RGB tip; they
are thus mostly AGB stars. Sources detected only at 7~$\mu$m are mainly normal
bulge red giants just below the RGB tip.

The completeness and reliability of detection at 7~$\mu$m and 15~$\mu$m
of point sources are high and well quantified down to $\sim~$15 mJy and
$\sim~$10 mJy respectively.  The photometric accuracy is reduced 
by a variety of effects: by the complex time-
and illumination history-dependant behaviour of ISOCAM pixels,
especially in regions with a very high density of bright sources, and
also by the integration time available for a wide-area survey. However,
the photometric accuracy we have achieved is good enough to be able to
take advantage of the rich information provided by the combination of
the five wavelength data of ISOGAL+DENIS. Of particular importance is
our ability to detect reliably quite small reddening-corrected
15~$\mu$m excesses. A detailed analysis of the stellar populations is
also much eased by the overwhelming preponderance of bulge or central disk 
stars with
a well defined distance, and, in the present field, by the relatively
low and constant interstellar reddening in front of the whole field
studied. We have shown here that our ISOGAL survey is ideal (though
not unique) for analysis of AGB stars with large mass-loss rates
($\dot M~\ge~10^{-7}M_{\odot}$/yr), by providing a complete census in the
field, independent of large amplitude variability.

The most important conclusion for future analyses from the present
analysis is our demonstration that the combination of near-IR (DENIS)
and mid-IR (7~$\mu$m and 15~$\mu$m) ISOGAL data allows reliable
detection of circumstellar dust and low rates of mass-loss in bulge
AGB stars not deducible from near-infrared data alone.  ISOGAL is
uniquely suitable for systematic studies of AGB stars with low rates
of mass-loss in the whole inner Galaxy, especially in the bulge.  The
very small amounts of dust associated with low rates of mass loss are
undetectable in the near-infrared, both in absorption and in emission,
while being readily detectable at 15~ $ \mu$m.
Stars with low rates of mass loss are
too faint to have been detectable by IRAS, except in the solar
neighbourhood; most of them will escape detection by MSX because of
the confusion limit arising from the 18$\arcsec$ MSX pixels, which have areas 
almost
an order of magnitude larger than those used for ISOGAL.
Although mid-infrared evidence for dust emission
corresponding to low rates of mass loss is seen in the
IRAS data for AGB stars in the solar neighbourhood, inevitable
distance uncertainties make the analysis of such IRAS data for
mass-loss rates and AGB evolution much less clear-cut.

The most important immediate scientific conclusion of this paper is
our detection of low rates of mass loss which are ubiquitous for red giants 
with luminosities just above or possibly close to the RGB tip, and thus still 
in 
relatively early stages.
The most luminous of these stars, 
that we defined as ``intermediate'' AGB (in the early AGB thermal pulse phase), 
form a well defined sequence
in the [15]/K$_s$-[15] and [15]/[7]-[15] magnitude-colour planes.  In
order to explain the colours the presence of dust is required, with
(model-dependant) dust mass-loss rates of a few
10$^{-11}~M_{\odot}$/yr.
It is thus well established that dust formation is already
associated with weak mass-loss during the early TP-AGB phase,
This obviously puts important constraints on the
physics of dust formation.

These first results can obviously be improved and exploited in several
ways. For the data presented for this specific field, we still hope to
improve the photometry and the reliability of the ISOGAL data, in
particular by including the verification observations not yet fully
exploited. The study of mass-losing red giants will be extended to the
other ~200 ISOGAL fields : i) in a straightforward way to the other
ISOGAL bulge fields with $ |b|~>~1^{0}$, as already done for the
ISOGAL observations of two Baade Window fields (Glass et
al. \cite{Glass99}); ii) to the bulk of the ISOGAL fields closer to
the Galactic plane, where there is large and variable extinction, and
where the uncertainty on the distance and the mixing with young stars
somewhat complicate the analysis.  We are currently analysing the
surface density of the various classes of AGB stars to characterize
the structure and stellar populations of the bulge, to determine to
what extent it is meaningful to consider the various structures:
bulge, central disk, bar, central cluster, and so on.

As concerns the theoretical interpretation, much work is still needed
in order to:

i) Improve the models of 
red giants with weak mass-loss both for photospheric
and for dust emission. Many questions can be addressed: what is the
chemical nature of the dust, considering silicates and possible other
components; is the base of the large spread we have observed in I-J
colours at fixed luminosity a metallicity effect; and more generally,
how can we disentangle the effects due to initial mass, age and metallicity?

ii) Use our ISOGAL results to further constrain the models of dust
formation and of TP-AGB evolution, and in particular: determine whether dust 
formation can begin in RGB stars close to the RGB tip, or whether it is 
specific of AGB stars; and explain 
dust formation in the context of very weak mass-loss.

{\it Acknowledgements}\\ 
This work was carried out in the context of EARA, the European Association  for
Research in Astronomy.   We would like to thank M. P\'erault, P. Hennebelle, 
S. Ott, R. Gastaud, H. Aussel, and F. Viallefond for useful
discussions during the course of the reduction of the ISO data. 
We thank the referees, especially, J.A. Frogel, for their very constructive 
and useful suggestions.
SG was supported by a fellowship 
from the Ministere des Affaires Etrang\`eres, France.  MS acknowledges the 
receipt of an ESA fellowship. The DENIS project is partially funded by European 
Commission through SCIENCE and Human Capital and Mobility plan grants. It is also 
supported, in France by the Institut National des Sciences de l'Univers, 
the Education Ministry and the Centre National de la Recherche Scientifique, 
in Germany by the State of Baden-W\"urtemberg, in Spain by the DG1CYT, in Italy by 
the Consiglio Nazionale delle Ricerche, in Austria by the Fonds zur F\"orderung 
der wissenschaftlichen Forschung und Bundesministerium f\"ur Wissenshaft und 
Forschung, in Brazil by the Foundation for the development of Scientific
Research of the State of Sao Paulo (FAPESP), and in Hungary by an OTKA grant
and an ESOC\&EE grant.

\appendix

\section{Definition of ISOGAL Fluxes and Magnitudes}

The fluxes and magnitudes used were defined in the following way
(Blommaert \cite{Blommaert98}). The ISOCAM units of ADU/gain/sec were
first converted into mJy/pixel units within CIA, with the conversion
factors:

\begin{equation}
F(mJy) = (ADU/G/s)/2.33
\label{eq:cnv1}
\end{equation}
for LW2 and

\begin{equation}
F(mJy) = (ADU/G/s)/1.97
\label{eq:cnv2}
\end{equation}
for LW3. These are correct for a F$_{\lambda}$ $\sim$ $\lambda$$^{-1}$ power 
spectrum at wavelengths 6.7~$\mu$m and 14.3~$\mu$m respectively.

Magnitudes are defined by 
\begin{equation}
mag(LW2) = [7] = 12.39 - 2.5 \times log[F_{LW2}(mJy)]
\label{eq:cnv3}
\end{equation}

\begin{equation}
mag(LW3) = [15] = 10.74 - 2.5 \times log[F_{LW3}(mJy)]
\label{eq:cnv4}
\end{equation}
where the zero point has been chosen to provide zero magnitude for a
Vega model flux (A0V star, not including the infrared excess emission of the
circumstellar disk) at the respective wavelengths mentioned earlier.

\section{Details of Data Reduction and Quality; Cross-Identifications}

\subsection{ISOGAL}

The special procedure for ISOGAL source extraction, developed by Alard et al (
in preparation), uses several ``CIA'' \footnote{``CIA'' is a joint development
by the ESA Astrophysics Division and the ISOCAM Consortium.}  procedures not
yet implemented in the standard (auto-analysis) treatment (corrections for
distortion of the ISOCAM field, and for time behaviour [``vision'' and `` 
inversion'' methods]). It also includes a sophisticated source
extraction, after a regularisation of the point-spread-function
(PSF). Indeed, there is not yet a good standard procedure to fully
correct for the time behaviour effects for fields such as ISOGAL ones
with strong sources and background.  Here we have used the fluxes
given by the ``inversion'' method, which provides better photometry,
but we have used the ``vision'' method to identify and drop the false
sources generated by the detector ``memory'' of bright sources
previously observed at a different location on the sky, but in the
same pixel.
 
After the elimination of the false replication sources, the source counts in
this field are 599 and 282 respectively in LW2 ([7] $ < $ 9.75) and LW3 ([15]
$ < $ 8.5) . This is indeed close to the confusion limit for LW2 sources (85
pixels [$3\arcsec \times 3\arcsec$] per source, density 1.7~10$^4$
deg$^{-2}$), but farther from this limit for LW3 ones (8.1~10$^3$~deg$^{-2}$).

\subsection{DENIS}
For DENIS sources, the sensitivity is mostly limited by confusion in the K$_s$
and J bands ($\sim 7.7 \times 10^4~deg^{-2}$ [for K$_s$ $<$ 12 and J $<$ 14],
giving $\sim$19~pixels $(3\arcsec \times 3\arcsec)$ per source). The
completeness limit is thus probably close to 11.5 in the K$_s$ band and 13.5
in the J band. The density at the sensitivity limit in the I band, $\sim18$ ,
is farther from confusion (density $\sim 1.3~10^5 deg^{-2}$ with $1\arcsec
\times 1\arcsec$ pixels). 

Independent magnitudes are available for many  DENIS sources in the
overlap region between adjacent observations.  Analysis of these
repeated observations shows that the internal dispersion in the
photometry, in this crowded region, is less than 0.1 mag for K$_s$ $<$ 11, J
$<$ 13.5 and 
I $<$ 16.5 (it rises to 0.18 mag for K$_s$ $<$ 13, 0.13 mag for
J $<$ 14.5 and 0.2 for I $<$ 17.5).  For the determination of the zero
point all standard stars observed in this night have been used. We
derived the following zero points: $\rm I = 23.45$, $\rm J = 21.59$
and $\rm K_{s} = 19.85$, respectively.  
The internal rms in the zero-points is found to be 
0.03, 0.07 and 0.04 mag in the K$_s$, J and I bands respectively.

\subsection{Cross-Identifications}

We have now routine standard
procedures for ISOGAL-ISOGAL and DENIS-ISOGAL cross-identifications
(Copet et al. in preparation). The good quality of the pointing of ISO
and of the correction of the ISOCAM field distortions permits, after
optimisation of a small rotation-translation of the fields, a
reduction of the rms of the nominal offsets of matched sources to
$\sim 0.6\arcsec$ and $\sim 1.1\arcsec$ for LW3/LW2 and ISOGAL/DENIS
respectively. However, the search radius was fixed at a large value,
2.7$\arcsec$, for LW3/LW2 associations in order not to miss
associations. The chance of spurious association with an LW2 source is
then $\sim 4\%$. Because of the very high density of DENIS sources,
the search radius was reduced to 2.1$\arcsec$ for the DENIS/ISOGAL
associations. Nevertheless, the density of the DENIS sources is so
high that the chance of spurious associations remains $\sim~10\%$ for
K$_s$ sources with K$_s$ $<$ 12. The chance of spurious association is
reduced to $5\%$ when one limits the associations to K$_s$ = 11.
A substantial fraction of the ISO sources have thus been identified with
DENIS sources. Out of a total number of 599 LW2 sources, 557 ($93\%$)
are matched with a K$_s$ $<$ 12 source, 552 with a JK$_s$ source and
522 with an IJK$_s$ source. Out of 282 LW3 sources, 248 ($86\%$) are
matched with an LW2 source, 237 ($84\%$) with a JK$_s$/LW2, and 221
(78$\%$) with an IJK$_s$/LW2 source. The number of LW2/LW3 sources
without K$_s$ or LW3/K$_s$ sources without LW2 is very small, 10 in both cases
.

\subsection{ISOGAL Completeness and Photometry}

In order to check the completeness of LW3 sources, we can use the more
sensitive LW2 observations. One can check, for instance, that only 13
LW2 sources with [7] $<$ 8.3, among 183 in total,
are missing in LW3. From the known range of values of
[7] - [15], we can conclude from Figure \ref{ISO_CMPL} that the
completeness of LW3 sources is close to $100\%$ for [15] $<$ 7.5
($\sim20$~mJy) and $\sim65-90\%$ in the range 7.5 $<$ [15] $<$ 8.5
($\sim8-20$~mJy).

DENIS K$_s$ detections can be used in a similar way to estimate the
completeness of LW2 sources. Figure \ref{ISO_CMPL} compares the number
of LW2 sources detected per half magnitude bin with an approximate estimate
of the number expected from K$_s$ sources with typical colours. It is seen that the
detections are practically complete for [7] $<$ 8.5 ($\sim 35$ mJy) and that 
they remain more than 80$\%$ complete for 8.5 $<$ [7] $<$ 9.5 ($\sim 15 - 35$ mJy);
however, the completeness rapidly decreases below $\sim~15$~mJy. This incompleteness 
is mainly due to confusion.

The quality of ISOGAL photometry has been checked in this field (Table~1) and
others by repeated observations  both with 
6$\arcsec$ pixels (Ganesh et al. in preparation). The uncertainty thus
proved to be better than $\sim 0.2$ mag rms above $\sim 15$ mJy in
both bands. It is poorer for weaker sources, especially in the LW3
band. One can expect a similar repeatibility accuracy with 3$\arcsec$
pixels. However, this does not take into account systematic errors. 
In particular, the comparison of the 3$\arcsec$ and 6$\arcsec$ pixels
measurements which were performed on this field shows a small
systematic difference in the fluxes, with average differences up to
0.1-0.2 mag rms. Further work is in progress to understand these details, but
the effect may be explained by the source confusion as is discussed in DePoy
et al (\cite{Depoy93}). Our photometry is thus still
uncertain by a few tenths of a magnitude systematically. 

The photometry is expected to be poorer on the edges of
the ISOGAL image: in such a small raster (4~$\times$~7 pointings),
$\sim 40\%$
the image is observed with a single exposure instead of the double
exposure on average for the points of the central part. In addition,
the source extraction is not able to recover the full intensity of
sources very close to the edges within a few pixels.


\end{document}